\documentclass[prc,twocolumn,showpacs,amsmath,amssymb]{revtex4}
\usepackage{times}
\usepackage{graphicx}
\usepackage{dcolumn}
\usepackage{bm}
\usepackage{amstext}

\begin{document}

\title{The structural evolution in transitional nuclei of mass 82 
$\leq$ A $\leq$ 132}
\author{M. Bhuyan\footnote {Email: bunuphy@itp.ac.cn}}
\affiliation{State Key Laboratory of Theoretical Physics, Institute of 
Theoretical Physics, Chinese Academy of Sciences, Beijing 100190, China}  

\date{\today}

\begin{abstract}
In this theoretical study, we report an investigation on the behavior 
of two-neutron separation energy, a differential variation of the nucleon 
separation energy, the nuclear charge radii and the single-particle energy 
levels along the isotopic chains of transitional nuclei. We have used the 
relativistic mean field formalism with NL3 and NL3$^*$ forces for this 
present analysis. The study refers to the {\it even-even} nuclei such as 
Zr, Mo, Ru and Pd for N = 42$-$86, where a rich collective phenomena such 
as proton radioactivity, cluster or nucleus radioactivity, exotic shapes, 
{\it Island of Inversion} and etc. are observed. We found that there are 
few non-monotonic aspects over the isotopic chain, which are correlated 
with the structural properties like shell/sub-shell closures, the shape 
transition, clustering and magicity etc. In addition to these, we have 
shown the internal configuration of these nuclei to get a further insight 
into the reason for these discrepancies.
\end{abstract}


\maketitle

\section{Introduction}
Nowadays, one of the most sensitive and crucial region in the nuclear chart 
for investigation is laying in between $Z$= 35$-$64 and $A$= 82$-$132. This 
region reveals a large number of interesting discoveries of new phenomena, 
such as proton radioactivity \cite{karny12,blank08,olsen13}, cluster 
radioactivity \cite{rose84,gupta94,bhu12}, exotic shapes 
\cite{gaff13,lister13}, {\it Island of inversion}  \cite{thib75,tara09}, 
abnormal variation of major shell closures (i.e. extra stability near 
drip-line) \cite{ohn08,ohn10,wata13} and giant halo near neutron drip-line 
region \cite{ring98} etc. These crucial features may be due to the rapid 
growing possibility of the neutron-proton ratio ($N/Z$) in a nucleus. From 
last few decades, it is possible to study these exotic nuclei by using the 
radioactive isotope beams (RIB) facilities. This reveals the new concept 
entitled as {\it aforementioned} magic number. In other word, the confirmation 
of magic number near $\beta-$stability line are not mandatory universal 
\cite{iwas00,moto95,bast07}. Further, the structural properties of nuclei 
far away from the $\beta-$stability line are also active areas of research 
in both theories and experiments \cite{ring98,wata13,mayer49}. In particular, 
the neutron-rich Zr$-$, Mo$-$, Ru$-$ and Pd$-$ with mass numbers $A$ 
=100$-$130 are of special interest for various reasons. For example, they 
lie far away from the $\beta-$stable region of {\it Nuclear Landscape}, result 
in a well established deformation, but close enough in the magnitude of 
microscopic excitations to compete with the collectivity of double shell 
closure nuclei \cite{ring98,blon89,sumi11}. Moreover, these nuclei are also 
holding an active participation in the nucleosynthesis of heavy nuclei in 
astrophysical {\it r-}process. The mass and decay properties are quite an 
essential ingredient to building up the path, the isotopic abundances and 
the time period of these process \cite{krat05}.

In addition to that the nuclear structure of these nuclei is characterized 
by a strong competition between various shapes, which gives rise to the 
shape instabilities that lead to coexistence nuclear shape transitions in 
the isotopic chains \cite{nish12}. This could be understood from the potential 
energy surface at different deformations. Elaborately, the occurrence of two 
(or more) nearly equally deep minima in the potential energy surface at 
different deformations show the signature for nuclear shape coexistence. 
Hence, one can say the nuclear shape are not only vary with the nucleon 
number but also with the excitation energy and spin. It is well known that 
the binding energy of a nucleus is one of the most precise measured 
observable from the experiments \cite{tu11,zhang12}. Several nuclear 
observables which are highly relevant for understanding various features 
of the nuclear structure can be computed from its mass such as the average 
nuclear field, nucleon-nucleon (NN) potential, single particle energy etc. 
The correlations among these fundamental quantities are amended to explain 
the deformed ground states, low-lying isomeric states and few derived 
quantities like moments of inertia and vibrational excitation energy etc 
\cite{myer70,myer83,tanihata85,lass92}. It is acclaimed that the energy 
involved in removal of Fermions from a strongly correlated system of identical 
Fermions must be a good indicator for the stability of the system. This 
magnitude of this energy has much higher values for systems with {\it even} 
number of particles than {\it odd} one, if the pairing is a dominant component 
in the binary {\it Fermion$-$Fermion} interaction.

In this present work, the quantities of interest are the nuclear potential 
energy surface, nuclear shape, nuclear binding energy, two neutron separation 
energies ($S_{2n}$), the differential variation of neutron separation energy 
$\Delta S_{2n}$, the root-mean-square charge distribution $r_\mathrm{ch}$ 
and the single particle energy level for the {\it even-even} mass transition 
nuclei. Base on these decisive observables, we have focused on the evolution 
on the structural properties of transition nuclei. The paper is organized as 
follows: Section II gives a brief description of the relativistic mean field 
formalism including the pairing energy correlation. The results of our 
calculation along with discussions are presented in Section III. Section IV 
includes a short summary along with few concluding remarks.

\section{The relativistic mean-field (RMF) method}
From last few decades, the nuclear covariant density functional
theories (CDFT) are quite successful in describing the ground and the
intrinsic excited state including fission states of the exotic heavy
and superheavy nuclei over the nuclear chart
\cite{bur04,blum94,bend03,lu06,abu10,lu12,lala12,pras13}. Basically,
there are four different patterns to perform the  covariant density
functional: the point coupling nucleon or meson exchange  interactions
connected with the density-dependent or nonlinear couplings. One can
also introduced all shape degrees of freedom to CDFTs by breaking both
the axial and reflection symmetries simultaneously (see the Ref.
\cite{lu12} for more details). In the relativistic mean field approach,
the nucleus is considered as a composite system of nucleons (proton and
neutron) interacting through exchanges of mesons and photons
\cite{bend03,sero86,rein89,ring96a,vret05,meng06,paar07}. Further, the
contributions from the meson fields are described either by mean fields
or by point-like interactions between the nucleons \cite{niko92,bur02}.
and the density dependent coupling constants \cite{fuch95,niks02} or 
nonlinear coupling terms \cite{bogu77,bro92} are introduced to reproduced 
the correct saturation properties of infinite nuclear matter. Here, most
of the computational efforts are devoted to solving the Dirac equation
and to calculate various densities. In the present calculation, we have
used the microscopic self-consistent relativistic mean field (RMF) theory
as a standard tool to investigate the nuclear structure phenomena. It is
worth mentioning that the RMF approach is one of the most popular and
widely used formalism among them. The relativistic Lagrangian density
(after several modification of the original Walecka Lagrangian to take
care of various limitations) for a nucleon-meson many body system
\cite{ring86,sero86,lala99c,bhu09,bhu11,rein89,ring96a,vret05,meng06,paar07,niks11}
is given as:
\begin{eqnarray}
{\cal L}&=&\overline{\psi_{i}}\{i\gamma^{\mu}
\partial_{\mu}-M\}\psi_{i}
+{\frac12}\partial^{\mu}\sigma\partial_{\mu}\sigma
-{\frac12}m_{\sigma}^{2}\sigma^{2}\nonumber\\
&& -{\frac13}g_{2}\sigma^{3} -{\frac14}g_{3}\sigma^{4}
-g_{s}\overline{\psi_{i}}\psi_{i}\sigma-{\frac14}\Omega^{\mu\nu}
\Omega_{\mu\nu}\nonumber\\
&&+{\frac12}m_{w}^{2}V^{\mu}V_{\mu}
-g_{w}\overline\psi_{i}\gamma^{\mu}\psi_{i} V_{\mu}
-{\frac14}\vec{B}^{\mu\nu}.\vec{B}_{\mu\nu}\nonumber \\
&&+{\frac12}m_{\rho}^{2}{\vec
R^{\mu}} .{\vec{R}_{\mu}}
-g_{\rho}\overline\psi_{i}\gamma^{\mu}\vec{\tau}\psi_{i}.\vec
{R^{\mu}}\nonumber\\
&&-{\frac14}F^{\mu\nu}F_{\mu\nu}-e\overline\psi_{i}
\gamma^{\mu}\frac{\left(1-\tau_{3i}\right)}{2}\psi_{i}A_{\mu} .
\end{eqnarray}
From the above Lagrangian, we obtain the field equations for the nucleons and
mesons. These equations are solved by expanding the upper and lower components
of the Dirac spinors and the boson fields in an axially deformed harmonic
oscillator basis, with an initial deformation $\beta_{0}$. The set of coupled
equations are solved numerically by a self-consistent iteration method
\cite{horo81,boguta81,price87,fink89}. Based on the effective interactions
used in the RMF functional, the center of mass energy can be calculated
either in harmonic oscillator approximation or from the quasi-particle vacuum
self-consistently. In case of oscillator approximation, the spurious center
of mass motion is substracted using the Elliott-Skyrme approximation
\cite{elli55}. The analytical form is given as:
\begin{equation}
E_{\mathrm{c.m}} = \frac{3}{4}\time41A^{-\frac{1}{3}},
\end{equation}
where $A$ is the mass number. In other hand, one should estimate the center of
mass energy using self-consistent method \cite{suga94,long04}, 
\begin{equation}
E_{\mathrm{c.m}} = \frac{\langle F \vert P^2 \vert F \rangle}{2M}, 
\end{equation}
where, $\vert F \rangle = \vert F \rangle_{RMF}$ wave function. The $P$ and 
$A$ are the total linear momentum and the nuclear mass number, respectively.
The results obtained from these two methods for Zr isotopes are given in 
Table I. From the table it is clear that, the calculated center of mass 
energies from both the cases are almost overlap with each other. Hence, one 
can use any one of the method for the center of mass energy correction in the 
calculation of this nuclear mass region (see Ref. \cite{long04}). The total 
quadrupole deformation parameter $\beta_2$ is evaluated from the resulting 
proton and neutron quadrupole moments, as
\begin{eqnarray}
Q=Q_n+Q_p=\sqrt{\frac{16\pi}5} (\frac3{4\pi} AR^2\beta_2).
\end{eqnarray}
The root mean square (rms) matter radius is defined as
\begin{eqnarray}
\langle r_m^2\rangle=\frac{1}{A}\int\rho(r_{\perp},z) r^2d\tau,
\end{eqnarray}
where $A$ is the mass number, and $\rho(r_{\perp},z)$ is the axially deformed 
density. The total binding energy and other observables are also obtained by 
using the standard relations, given in \cite{ring86}. As outputs, we obtain 
different potentials, densities, single-particle energy levels, radii, 
deformations and the binding energies. For a given nucleus, the maximum 
binding energy corresponds to the ground state and other solutions are 
obtained as various excited intrinsic states.
\begin{table*}
\caption{The center-of-mass energy obtained from phenomenological 
\cite{elli55} and microscopic self-consistent methods \cite{suga94,long04} 
for Zr isotopes.}
\renewcommand{\tabcolsep}{0.17cm}
\renewcommand{\arraystretch}{2.4}
\begin{tabular}{ccccccccccccccccccc}
\hline\hline
Method&$^{100}$Zr&$^{102}$Zr&$^{104}$Zr&$^{106}$Zr&$^{108}$Zr&$^{110}$Zr&
$^{112}$Zr&$^{114}$Zr&$^{116}$Zr&$^{118}$Zr&$^{120}$Zr \\
\hline
$E_{\mathrm{c.m}}=\frac{3}{4}\time41A^{-\frac{1}{3}}$ & $-$6.63 & $-$6.58 & $-$6.54 
& $-$6.46 & $-$6.46 & $-$6.42 & $-$6.38 & $-$6.34 & $-$6.31 & $-$6.27 & $-$6.23 \\
\hline
$E_{\mathrm{c.m}}=\frac{\langle F \vert P^2 \vert F\rangle}{2M}$ & $-$6.83 & $-$6.79 
& $-$6.75 & $-$6.68 & $-$6.61 & $-$6.55 & $-$6.49 & $-$6.45 & $-$6.41 & $-$6.44 & $-$6.47 \\
\hline\hline
\end{tabular}
\label{Table 1}
\end{table*}
%%%%%%%%%%%%%%%%%%%%%%%%%%%%
\begin{figure}
%\vspace{0.6cm}
\begin{center}
\includegraphics[width=1.0\columnwidth]{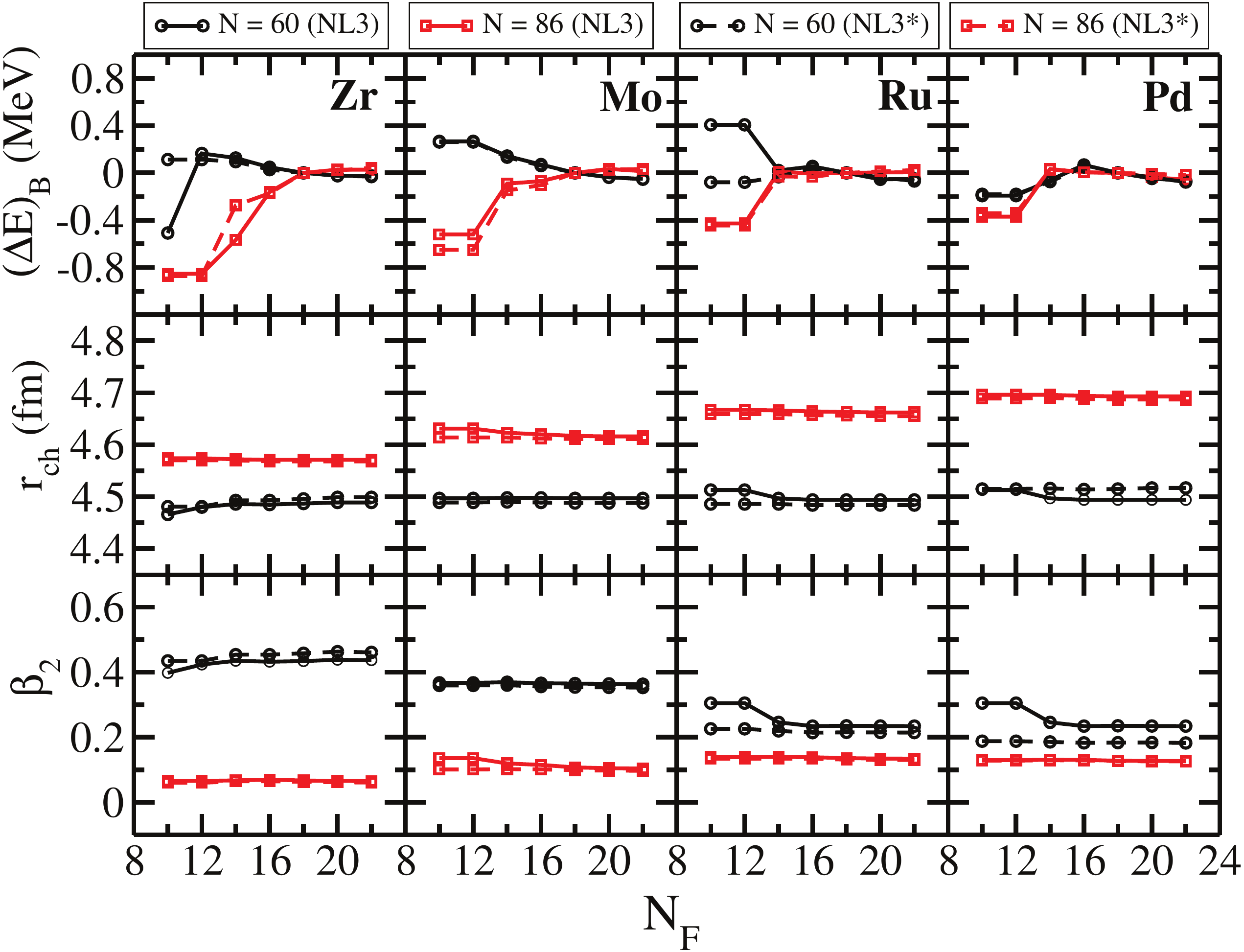}
\caption{(Color online) The obtained results for nuclear relative binding 
energy difference $(\Delta E)_B = E_B (N_F=18)- E_B (N_F=10-22)$, charge 
radius $r_{ch}$ and the quadruple deformation $\beta_2$ for $^{100,126}$Zr, 
$^{102,128}$Mo, $^{104,130}$Ru and $^{106,132}$Pd isotopes from NL3 and NL3* 
force parameter. See text for more details.}
\end{center}
\label{Fig. 1}
\end{figure}
%%%%%%%%%%%%%%%%%%%%%%%%%%%%%%%%%%%%%%%%%%%
\subsection{Pairing Energy}
To deal the nuclear bulk properties of open-shell nuclei, one can not 
neglect the pairing correlation in their ground and intrinsic excited state 
\cite{karat10}. There are various methods such as the BCS approach, the 
Bogoliubov transformation and the particle number conserving methods have 
been developed to treat the pairing effects in the study of nuclear 
properties including fission barriers \cite{zeng83,moli97,zhang11,hao12}. 
In principle, the Bogoliubov transformation is widely used methods to take 
pairing correlation into account for the drip-line region 
\cite{vret05,paar07,ring96a,meng06,lala99a,lala99b}. In case of nuclei not 
too far from the $\beta$-stability line, one can use the constant gap BCS 
pairing approach reasonably good to take care of pairing \cite{doba84}. 
Further, the BCS approach may fail for neutron-rich light nuclei. In the  
present analysis, we have considered the intermediate mass neutron-rich 
nuclei, hence the RMF results with BCS treatment should be reliable. In 
other word, to avoid the difficulty in the calculation, one can employ the 
constant gap BCS approach to deal the present mass region 
\cite{madland81,moller88,bhu09}. Now the expression for pairing energy is given by, 
\begin{equation}
E_{pair}=-G\left[\sum_{i>0}u_{i}v_{i}\right]^2
\end{equation}
where $v_i^2 + u_i^2 = 1$ is known as the occupation probability and $G$ 
is the pairing force constant \cite{ring90,patra93,map82}. The variational 
procedure with respect to the occupation numbers $v_i^2$, gives the BCS 
equation $2\epsilon_iu_iv_i-\triangle(u_i^2-v_i^2)=0$ and the pairing gap 
$\Delta$ is defined by 
\begin{equation}
\triangle=G\sum_{i>0}u_{i}v_{i}.
\end{equation}
This is the famous BCS equation for pairing energy. The densities are 
contained within the occupation number, 
\begin{equation}
n_i=v_i^2=\frac{1}{2}\left[1-\frac{\epsilon_i-\lambda}
{\sqrt{(\epsilon_i-\lambda)^2+\triangle^2}}\right].
\end{equation}
The standard expression for the pairing gaps of proton and neutron are 
$\triangle_p = RB_s e^{sI-tI^2}/Z^{1/3}$ and 
$\triangle_n = RB_s e^{-sI-tI^2}/A^{1/3}$, respectively \cite{madland81}. 
Here the constants and their values are as follows: $R$ = 5.72, $s$ = 0.118, 
$t$ = 8.12, $B_s$ = 1, and $I = (N-Z)/(N+Z)$. (Note that the gaps obtained 
by these expressions are valid for nuclei both on or away from the stability 
line for this mass region). The pairing force constant {\it G} is not 
calculated explicitly from the RMF equations. Using the above gap parameter, 
we calculate the occupation probability and the chemical potentials 
$\lambda_n$ and $\lambda_p$ from the particle  numbers using the above 
equations. Now, we can rewrite the pairing energy as,
\begin{equation}
E_{pair}= -\frac{\Delta^2}{G} =-\triangle\sum_{i>0}u_{i}v_{i}.
\end{equation}
Since it depends on the occupation probabilities $v_i^2$ and $u_i^2$, the 
pairing energy should change with particle number for a constant pairing 
gap. It is well known that the pairing energy $E_{pair}$ diverges if it is 
extended to an infinite configuration space for a constant pairing gap 
$\triangle$ and force constant $G$. Also, for the states spherical or 
deformed, with large momenta near the Fermi surface, $\triangle$ decreases 
in all the realistic calculations with finite range forces. However, for 
the sake of simplicity of the calculation, we have assumed, the pairing gap 
for all states is equal near the Fermi surface. In the present calculations 
we have used a pairing window, and all the equations extended up to the 
level $\epsilon_i-\lambda\leq 2(41A^{1/3})$, where a factor of {\it 2} has 
been included in order to reproduce the pairing correlation energy for 
neutrons in $^{118}$Sn using Gogny force \cite{ring90,patra93}. This kind 
of approach has already been used by many other authors in RMF and 
Skyrme-Hartree-Fock (SHF) models \cite{bhu09,ring90,patra93,bhu11}.
%%%%%%%%%%%%%%%%%%%%%%%%%%%%%%%%%%%%%%%%%%%%%%%%%%%%
\begin{table*}
\caption{The binding energy $E_B$, root-mean-square charge radius $r_{ch}$ 
and the quadrupole deformation parameter $\beta_2$ for the ground states 
and few selective first intrinsic ecxited state of $^{82-126}$Zr and 
$^{86-130}$Ru nuclei compare with the experimental data \cite{audi12}, 
wherever available. See the text for more details.}
\renewcommand{\arraystretch}{1.4}
\begin{tabular}{cccccccccc|ccccccccc}
\hline\hline
N&\multicolumn{3}{c}{RMF (NL3)}&\multicolumn{3}{c}{RMF (NL3*)}
&\multicolumn{3}{c|}{Experiment} & \multicolumn{3}{c}{RMF (NL3)}
&\multicolumn{3}{c}{RMF (NL3*)}&\multicolumn{3}{c}{Experiment}\\
\hline
& BE & $r_{ch}$ & $\beta_2$ & BE & $r_{ch}$ & $\beta_2$
& BE & $r_{ch}$ & $\beta_2$ & BE & $r_{ch}$ & $\beta_2$
& BE & $r_{ch}$ & $\beta_2$ & BE & $r_{ch}$ & $\beta_2$\\
\hline
\multicolumn{10}{c|}{{\bf $^{82-126}$Zr}} &\multicolumn{9}{c}{{\bf $^{86-130}$Ru}} \\
\hline \hline
42&691.3&4.29&$-$0.197&690.9&4.29&$-$0.192&     &      &0.367 &698.7 &4.39&$-$0.204&698.1 &4.39&$-$0.199&     &       &      \\
  &691.0&4.28&0.488   &690.6&4.28&0.488   &     &      &      &      &    &        &      &    &        &     &       &      \\
44&715.7&4.29&$-$0.188&715.8&4.27&$-$0.001&     &      &0.251 &726.8 &4.39&0.053   &726.8 &4.38&0.058   &     &       &      \\
  &715.5&4.28&0.469   &715.5&4.28&0.467   &     &      &      &      &    &        &      &    &        &     &       &      \\
46&739.2&4.28&0.001   &739.2&4.27&0.001   &740.6&      &0.151 &755.7 &4.39&0.096   &755.5 &4.38&0.089   &     &       &      \\
48&762.6&4.28&0.001   &761.9&4.28&0.001   &762.6&4.2812&0.185 &782.3 &4.37&0.006   &782.0 &4.37&0.004   &     &       &      \\
50&783.9&4.28&0.001   &783.1&4.28&0.000   &783.9&4.2696&0.089 &808.0 &4.37&0.001   &807.4 &4.37&0.000   &806.9&       &      \\
52&797.8&4.29&0.001   &797.2&4.28&0.000   &799.8&4.3057&0.1027&825.6 &4.39&0.003   &825.1 &4.38&0.002   &826.5& 4.3927&0.1579\\
54&810.5&4.34&0.169   &808.9&4.31&0.002   &814.7&4.3312&0.090 &843.2 &4.42&0.159   &842.5 &4.42&0.155   &844.8& 4.4232&0.1947\\
56&824.5&4.38&0.243   &822.9&4.38&0.233   &828.9&4.3498&0.080 &860.1 &4.45&0.205   &859.2 &4.44&0.199   &861.9& 4.4536&0.2148\\
58&837.0&4.42&0.318   &834.9&4.40&0.274   &840.9&4.4185&      &875.3 &4.47&0.225   &874.3 &4.47&0.216   &877.9& 4.4818&0.2404\\
60&849.8&4.48&0.432   &847.6&4.49&0.453   &852.2&4.5220&0.355 &889.3 &4.49&0.234   &888.2 &4.48&0.215   &893.0& 4.5104&0.2707\\
62&860.6&4.50&0.428   &858.2&4.50&0.428   &863.7&4.5690&0.427 &903.9 &4.57&0.385   &901.7 &4.52&0.295   &907.5&       &0.257 \\
64&870.6&4.52&0.427   &868.0&4.52&0.424   &873.8&      &0.38  &917.6 &4.54&$-$0.236&916.3 &4.53&$-$0.232&920.9&       &0.292 \\
  &     &    &        &     &    &        &     &      &      &916.9 &4.58&0.373   &915.8 &4.57&0.371   &     &       &      \\
66&880.4&4.54&0.419   &877.6&4.54&0.418   &883.9&      &      &931.0 &4.55&$-$0.236&929.5 &4.55&$-$0.234&933.3&       &0.295 \\
  &     &    &        &     &    &        &     &      &      &929.1 &4.59&0.357   &928.0 &4.59&0.357   &     &       &      \\
68&889.8&4.56&0.416   &886.8&4.56&0.419   &892.6&      &      &943.4 &4.57&$-$0.239&941.5 &4.57&$-$0.238&945.0&       &0.306 \\
  &     &    &        &     &    &        &     &      &      &941.5 &4.60&0.349   &940.1 &4.60&0.348   &     &       &      \\
70&897.2&4.59&0.445   &893.9&4.59&0.461   &900.4&      &      &954.1 &4.59&$-$0.233&951.8 &4.58&$-$0.229&     &       &      \\
72&904.1&4.62&0.478   &900.4&4.62&0.479   &     &      &      &963.9 &4.59&$-$0.203&961.5 &4.59&$-$0.196&     &       &      \\
74&911.8&4.52&-0.170  &908.8&4.52&-0.166  &     &      &      &974.2 &4.61&$-$0.178&971.5 &4.60&$-$0.176&     &       &      \\
76&917.7&4.52&-0.109  &914.5&4.52&-0.095  &     &      &      &983.1 &4.62&$-$0.169&979.9 &4.61&$-$0.167&     &       &      \\
78&923.4&4.52&0.065   &920.0&4.52&0.043   &     &      &      &991.6 &4.61&0.114   &988.7 &4.60&0.111   &     &       &      \\
80&929.6&4.54&0.003   &925.7&4.53&0.002   &     &      &      &999.5 &4.61&0.074   &996.2 &4.61&0.074   &     &       &      \\
82&935.6&4.55&0.001   &932.0&4.56&0.001   &     &      &      &1007.3&4.62&0.001   &1003.4&4.61&0.001   &     &       &      \\
84&936.7&4.56&0.001   &932.8&4.57&0.064   &     &      &      &1010.1&4.64&0.067   &1006.3&4.64&0.076   &     &       &      \\
86&937.3&4.57&0.069   &933.6&4.58&0.079   &     &      &      &1013.4&4.66&0.139   &1009.4&4.66&0.138   &     &       &      \\
\hline \hline
\end{tabular}
\label{Table 1}
\end{table*}
%%%%%%%%%%%%%%%%%%%
%%%%%%%%%%%%%%%%%%%
\begin{table*}
\caption{Same as Table II, only for $^{84-128}$Mo and $^{88-132}$Pd isotopes.
}
\renewcommand{\arraystretch}{1.4}
\begin{tabular}{cccccccccc|ccccccccc}
\hline\hline
N&\multicolumn{3}{c}{RMF (NL3)}&\multicolumn{3}{c}{RMF (NL3*)}
&\multicolumn{3}{c|}{Experiment} & \multicolumn{3}{c}{RMF (NL3)}
&\multicolumn{3}{c}{RMF (NL3*)}&\multicolumn{3}{c}{Experiment}\\
\hline
& BE & $r_{ch}$ & $\beta_2$ & BE & $r_{ch}$ & $\beta_2$
& BE & $r_{ch}$ & $\beta_2$ & BE & $r_{ch}$ & $\beta_2$
& BE & $r_{ch}$ & $\beta_2$ & BE & $r_{ch}$ & $\beta_2$\\
\hline
\multicolumn{10}{c|}{{\bf $^{84-128}$Mo}} &\multicolumn{9}{c}{{\bf $^{88-132}$Pd}} \\
\hline \hline
42&696.7&4.34&$-$0.206&696.2&4.34&$-$0.203&     &      &      &697.8 &4.42&0.002   &697.8 &4.42&0.002   &      &      &     \\
44&722.4&4.34&0.001   &722.6&4.34&0.002   &725.8&      &      &730.0 &4.44&0.094   &729.9 &4.44&0.095   &   &      &     \\
46&748.2&4.33&0.003   &748.2&4.33&0.003   &750.1&      &      &760.9 &4.43&0.101   &760.7 &4.43&0.104   &   &      &     \\
48&773.4&4.33&0.001   &773.1&4.33&0.001   &773.7&      &      &789.5 &4.42&0.004   &789.1 &4.42&0.005   &   &      &     \\
50&796.9&4.33&0.001   &796.4&4.33&0.001   &796.5&4.3156&0.1058&817.4 &4.42&0.001   &816.8 &4.41&0.001   &815.0&      &     \\
52&812.7&4.34&0.001   &812.2&4.34&0.001   &814.2&4.3518&0.1509&836.9 &4.43&0.003   &836.3 &4.42&0.004   &836.3&      &     \\
54&828.1&4.38&0.174   &827.1&4.37&0.158   &830.8&4.3841&0.1720&855.9 &4.46&0.136   &855.4 &4.46&0.139   &856.4&      &     \\
56&843.5&4.42&0.230   &842.2&4.41&0.220   &846.2&4.4088&0.1683&874.2 &4.48&0.176   &873.6 &4.48&0.177   &875.3&4.4839&0.196\\
58&857.2&4.45&0.268   &855.7&4.43&0.246   &860.5&4.4458&0.2309&891.1 &4.51&0.189   &890.5 &4.50&0.188   &892.8&4.5086&0.209\\
60&871.2&4.50&0.366   &869.1&4.49&0.356   &873.9&      &0.311 &906.9 &4.52&0.187   &906.2 &4.52&0.184   &909.5&4.5322&0.229\\
62&883.6&4.53&0.386   &881.4&4.52&0.382   &886.9&      &0.362 &921.8 &4.53&0.190   &921.1 &4.53&0.179   &925.2&4.5563&0.243\\
64&895.4&4.49&$-$0.234&894.0&4.49&$-$0.228&898.9&      &0.354 &936.1 &4.57&0.240   &934.9 &4.54&0.165   &940.2&4.5776&0.257\\
  &895.2&4.54&0.379   &893.9&4.54&0.377   &     &      &      &      &    &        &      &    &        &     &      &     \\
66&907.4&4.51&$-$0.236&905.7&4.51&$-$0.233&     &      &0.38  &951.8 &4.59&$-$0.231&950.4 &4.58&$-$0.229&954.3&      &0.220\\
  &906.5&4.56&0.374   &904.9&4.55&0.373   &     &      &      &950.3 &4.60&0.292   &949.5 &4.60&0.290   &     &      &     \\
68&918.1&4.53&$-$0.241&915.9&4.53&$-$0.239&     &      &      &965.7 &4.62&$-$0.23 &963.9 &4.60&$-$0.234&967.6&      &0.164\\
  &     &    &        &     &    &        &     &      &      &963.5 &4.62&0.304   &961.9 &4.60&0.301   &      &      &     \\
70&927.0&4.55&$-$0.231&924.6&4.54&$-$0.223&     &      &      &977.9 &4.63&$-$0.221&975.8 &4.61&$-$0.226&     &      &0.207\\
  &     &    &        &     &    &        &     &      &      &975.3 &4.60&0.216   &974.1 &4.59&0.216   &       &      &     \\
72&935.7&4.55&$-$0.197&933.2&4.55&$-$0.190&     &      &      &989.1 &4.63&$-$0.198&986.8 &4.62&$-$0.184&     &      &     \\
74&944.6&4.57&$-$0.180&941.7&4.56&$-$0.179&     &      &      &1000.6&4.63&$-$0.163&998.1 &4.62&$-$0.151&     &      &     \\
76&951.8&4.58&$-$0.172&948.4&4.58&$-$0.171&     &      &      &1011.9&4.63&0.115   &1009.8&4.62&0.114   &       &      &     \\
78&958.4&4.57&0.114   &955.0&4.56&0.094   &     &      &      &1022.8&4.64&0.104   &1020.2&4.63&0.105   &     &      &     \\
80&965.1&4.58&0.042   &961.5&4.57&0.037   &     &      &      &1032.4&4.65&0.073   &1029.3&4.64&0.076   &       &      &     \\
82&972.2&4.59&0.002   &968.0&4.59&0.001   &     &      &      &1041.5&4.65&0.001   &1037.8&4.65&0.002   &       &      &     \\
84&973.7&4.60&0.002   &969.6&4.60&0.029   &     &      &      &1044.5&4.66&0.027   &1040.9&4.67&0.043   &       &      &     \\
86&975.8&4.62&0.114   &971.7&4.61&0.101   &     &      &      &1048.7&4.69&0.130   &1045.1&4.69&0.132   &       &      &     \\
\hline \hline
\end{tabular}
\label{Table 1}
\end{table*}
%%%%%%%%%%%%%%%%%%%%%%%%%%%%%%%%%%%%%%%%%%%%%%%%%%%%%%%%%%%%%%
\section{Details of calculation and Results discussion}
In the present work, we have used the successful NL3 \cite{lala97} 
and the recently proposed NL3* \cite{lala09} force parameters, which 
are excellent in the description of ground and excited states with many 
collective aspect for spherical and deformed nuclei. In the mean time, 
there are several other mean-field interactions have been developed. 
In particular, the density dependent meson-exchange DD-ME1 \cite{niks02} 
and DD-ME2 \cite{lala05} interactions, which are adjusted to improve the 
isovector channel. Further, the density dependent points coupling 
interaction \cite{lu12,niks08} has been developed to describe the deformed 
heavy and superheavy nuclei. Even these interactions have been developed 
to provide a very successful description of various special features. At 
present, the $NL3$ \cite{lala97} and $NL3*$ \cite{lala09} forces are also 
accepted in compete with these parameters to reproduce the properties 
of the stable and nuclei far from the  $\beta$-stability line. In RMFT, 
the mean-field equations are solved self-consistently by taking different 
inputs of the initial deformation called $\beta_0$ 
\cite{lala97,lala09,ring86,ring90,bhu09}. To verify the convergence of 
the ground state solutions for this mass region, we pursued the calculation 
for $N_B$ =20 and varying $N_F$ from 10 to 22. The difference between the 
binding energy obtained  from $N_F$=18 to $N_F$=10-22 is entitled as relative 
binding energy difference and denoted as $(\Delta E)_B$. The estimated 
relative binding energy difference $(\Delta E)_B = E_B (N_F=18)- E_B 
(N_F=10-22)$, the charge radius $r_{ch}$ and the quadruple deformation 
$\beta_2$ for $^{100,126}$Zr, $^{102,128}$Mo,$^{104,130}$Ru and 
$^{106,132}$Pd isotopes from NL3 and NL3* force are shown in Fig. 1. From 
the figure, it is clear that the variations of these solutions are $\leq$ 
0.02$\%$ on binding energy and 0.01$\%$ on nuclear radii over the range 
of major shell Fermions $N_F$ from 10 to 14. But this realative changes 
are reduced to $\leq$ 0.002$\%$ on binding energy and 0.001$\%$ on nuclear 
radii for $N_F$ value from 14 to 22. Hence, the desired number of major 
shells for Fermions and bosons are $N_F$ = 18 and $N_B$ = 20 for the 
considered mass region. However, the number of mesh points for Gauss-Hermite 
and Gauss-Lagurre integration are $20$ and $24$, respectively. For a given 
nucleus, the solution corresponding to maximum binding energy is treated as 
a ground state and other solutions are the intrinsic excited states of the 
nucleus. 
%%%%%%%%%%%%%%%%%%%%%%%%%%%%%%%%%%%%%%
\begin{figure}
%\vspace{0.6cm}
\begin{center}
\includegraphics[width=0.9\columnwidth]{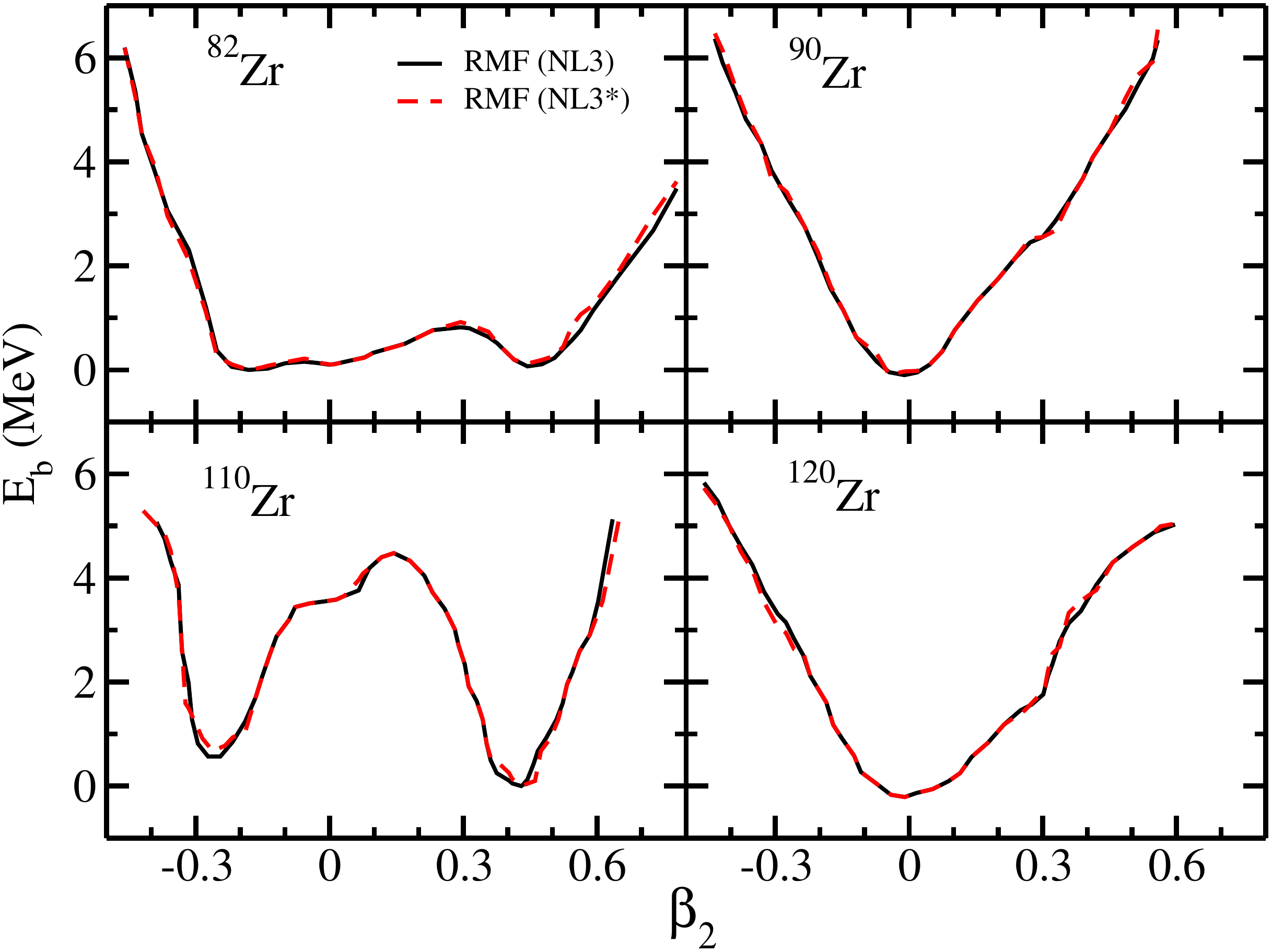}
\caption{(Color online) The potential energy surface of $^{82,90,110,120}$Zr
as a function of quadrupole deformation parameter $\beta_2$ for both $NL3$
and $NL3^*$ forces in axially deformed relativistic mean field calculations.
}
\end{center}
\label{Fig. 1}
\end{figure}
%%%%%%%%%%%%%%%%%%%%%%%%%%%%%%%%%%%%%%%%%

\subsection{Potential energy surface}
Conventionally, in case of a quantum mechanical system, the path followed by 
the different solutions at various deformation define a potential barrier or 
potential energy surface, which can be used for the determination of the ground 
state of a nucleus. More elaborately, from the potential energy surface (PES) 
obtained from a self-consistent relativistic mean field theory, one can 
regulate the reasonable results for the ground state similar to the 
non-relativistic calculations \cite{dubray08}. Since quadrupole deformation 
plays the most important and dominant part, we have neglected the other 
deformation coordinates in the present study for simplicity and low computation 
time cost. Here, the potential energy curve is calculated microscopically by the 
constrained RMF theory \cite{bhu09,bhu11,reinhard89,hirata88,zhou14}. The 
expectation value of the Hamiltonian \cite{ring90,hirata88,karat10} at certain 
deformation is given as,
\begin{equation}
H'=\sum_{ij}\frac{\langle\psi_i|H_0-\lambda Q_2|\psi_j\rangle}{<\psi_i|\psi_j>},
\end{equation}
where $\lambda$ is the constraint multiplier and $H_0$ is the Dirac mean 
field Hamiltonian. The convergence of the numerical solutions on the binding 
energy and the deformation are not very much sensitive to the deformation 
parameter $\beta_0$ of the harmonic oscillator basis for the considered range 
due to the large basis. Thus the deformation parameter $\beta_0$ of the 
harmonic oscillator basis is chosen near the expected deformation to obtain 
high accuracy and less computation time period.

The potential energy surface as a function of deformation parameter 
$\beta_2$, for the proton rich nucleus $^{82}$Zr, the double magic nucleus 
$^{90}$Zr and the neutron rich nucleus $^{110,120}$Zr are shown in Fig. 2, 
as a representative case. All other Mo$-$, Ru$-$ and Pd$-$ isotopes are also 
showing the similar behaviors, which are not given here. The energy ($E_b$ 
= $E_{g.s}-E_{e.s}$ on the $Y-$axis is the difference between the ground state 
energy to other constraint energy solutions. The solid and dotted line in the 
figure are for $NL3$ and $NL3^*$ force, respectively. The calculated $PES$ 
for both the cases are shown for a wide range from oblate to prolate 
deformations. We notice from the figure that there are more than one minima 
appear at different $\beta_2$. The magnitude of binding energy for the 
corresponding minima shows that the ground state solution appear at a certain 
value of $\beta_2$. The $\beta_2$ for the ground state is not same for all 
isotopes of Zr (see Table. II-III. For example, the ground state solutions 
for $^{82}$Zr, $^{90}$Zr, $^{110,120}$Zr and $^{120}$Zr are $\sim$ -0.2, 0.0, 
0.4 and 0.0, respectively. One can find similar nature for both the force 
parameter, hence one can conclude that the ground state properties of these 
nuclei are independent of the force parameters used.

\subsection{Nuclear Binding energy and quadrupole deformation}
The calculations mainly explain the nuclear structure as well as the 
sub-structure properties, based on the basic ingredients such as binding 
energy ($E_\mathrm{B}$), quadrupole moment $Q_{20}$, nucleonic density 
distribution $\rho (r_{\perp},z) = \rho_p (r_{\perp},z)+\rho_n(r_{\perp},z)$, 
and {\it rms} nuclear radii etc. Nevertheless, the present study demonstrates 
the applicability of RMF on the nuclear structure study for transition nuclei 
near neutron drip-line. The obtained results for binding energy $BE$, the 
quadrupole deformation parameter $\beta_2$ and the charge radius $r_{ch}$ for 
NL3 and NL3* force parameter for the isotopic chain of Zr, Mo, Ru and Pd are 
listed in Table II-III along with the experimental data \cite{audi12}. It is 
worth mentioning that the obtained results from the NL3 force parameter almost 
matches to the mass table by G. A. Lalazissis {\it et al.} \cite{lala99c} 
except few nuclei, though the small difference are acceptable in the accuracy 
of mean field level. Further, we notice on the binding energy and the 
$rms$ $r_{ch}$ for all nuclei over the isotopic chain from RMF agree well 
with the experimental values. Quantitatively, the mean deviation of $BE$ and 
$r_{ch}$ between the calculated result and the available experimental data 
over the isotopic chain are $\sim 0.01$ and $0.004$, respectively. Further, 
the quadrupole deformation parameter $\beta_2$, for both ground ($g.s.$) and 
selective excited states ($e.s.$) are also given in Table II-III. In some of 
the earlier RMF and Skyrme Hartree-Fock (SHF) calculations, it was shown that 
the quadrupole moment obtained from these theories reproduce the experimental 
data pretty well 
\cite{sero86,lala97,lala99a,bhu09,bhu11,ring90,brown98,patra1,patra2}. From 
the table, one can find that the shape of few nuclei is not consistent with 
the experimental observed shape. In this context, we have also estimated the 
first excited state solution for these nuclei correspond to the experimental 
deformations (see Table II-III). A careful inspection to these solutions shows 
that the small difference in the binding energy is an indication of shape 
coexistence. In other words, the two solutions in these nuclei are almost 
degenerate and might have large shape fluctuations. For example, in $^{82}$Zr 
the two solutions for $\beta_2$ = -0.197 and $\beta_2$ = 0.25 are completely 
degenerate with binding energies of 691.3 and 691.0 MeV, respectively. Hence, 
the ground state can be changed to the excited state and vice verse by a small 
change in the input, like the pairing strength, etc., in the calculations. 
Similar behavior is also observed for few other nucleus in the present analysis 
are listed in Table II-III. Such phenomenon is known to exist in many other 
mass regions of the nuclear chart \cite{yosi94,yosi94a} .
%%%%%%%%%%%%%%%%%%%%%%%%%%%%%%%%%%%%%%%%%%%%%
\begin{figure}
%\vspace{0.6cm}
\begin{center}
\includegraphics[width=0.9\columnwidth]{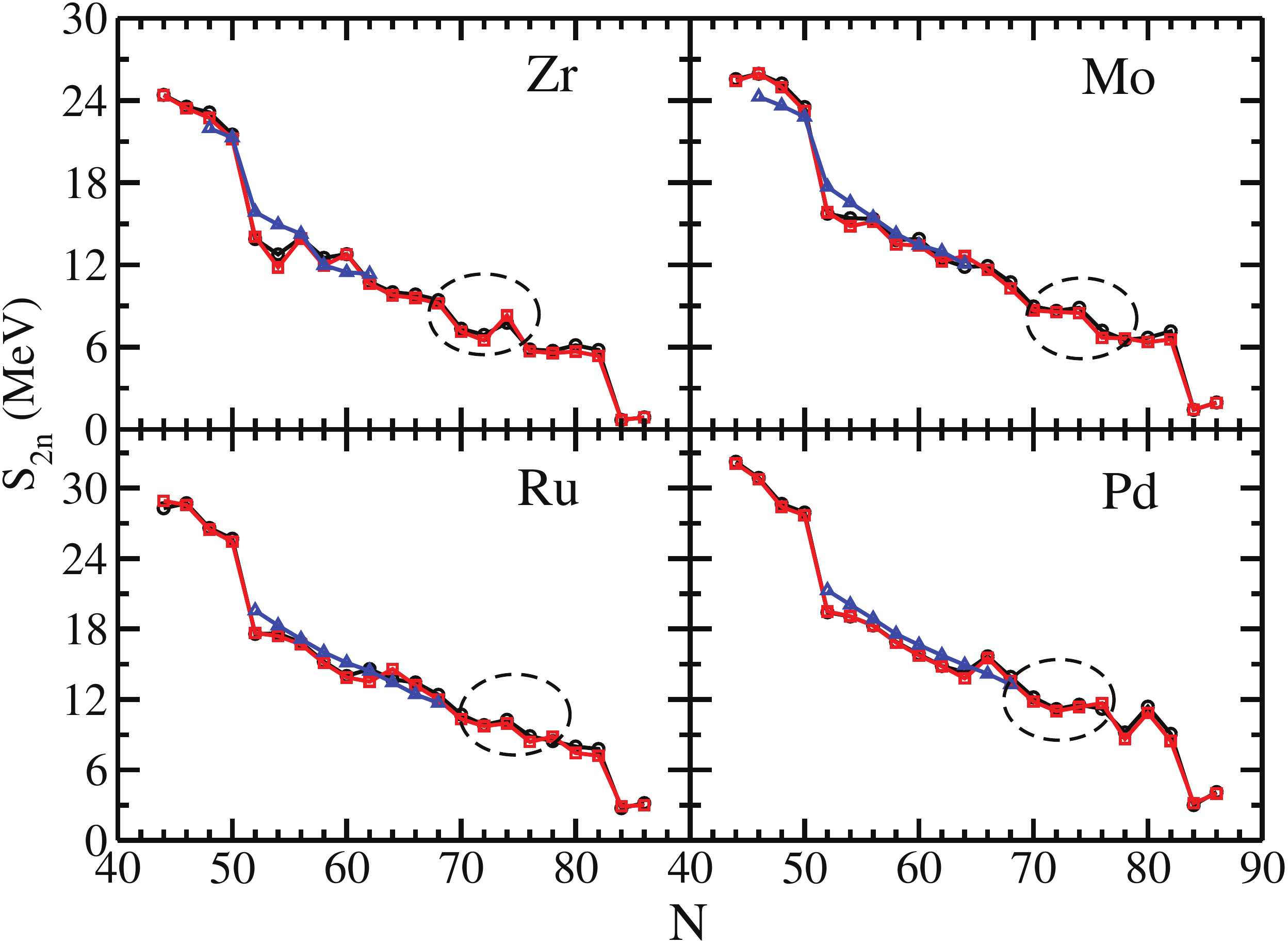}
\caption{(Color online) The two neutron separation energy as a function of
neutron number from RMF theory with $NL3$ and $NL3^*$ force parameter for
$^{82-126}$Zr, $^{84-128}$Mo, $^{86-130}$Ru and $^{88-132}$Pd nuclei are
compared with the experimental data \cite{audi12}.
}
\end{center}
\label{Fig. 1}
\end{figure}
%%%%%%%%%%%%%%%%%%%%%%%%%%%%%%%%%%%%%%%%%%

\subsection{Two neutron separation energy $S_{2n}$ (Z,N)}
Two neutron separation energy  $S_{2n}$ (Z, N), can be estimated from the ground 
state nuclear binding energies of $BE (Z, N)$, $BE (Z, N-2)$ and the neutron mass 
$m_n$ with the relation: 
\begin{eqnarray}
S_{2n} (Z, N) = -BE (Z, N) + BE (Z, N-2) + 2m_n,
\end{eqnarray}
The $BE$ of the $^AX_Z$ and $^{A-2}X_Z$ are calculated from RMF for NL3 and NL3* 
force parameters. It is essential to have very precise mass measurements to predict 
the correct estimation of the nucleon separation energy $S_{2n}$. The calculated 
$S_{2n}$ energy from RMF as a function of neutron number for Zr, Mo, Ru and 
Pd isotopes are compared with latest experimental data \cite{audi12}, shown in 
the Fig. 3. From the figure, it is clear that in an isotopic chain, the $S_{2n}$ 
energy shows the well-known regularities for a given atomic number i.e. the $S_{2n}$ 
decreases smoothly as the number of neutron increases in an isotopic chain. A sharp 
discontinuities (in other word kinks) appears at neutron magic numbers at $N$ = 50 
and 82. In energy terminology, one can write, the energy necessary to remove two 
neutrons from a nucleus (Z, $N_{magic}$+2) is much smaller than that to remove $two$ 
neutrons from the nucleus (Z, $N_{magic}$), which breaks the regular trend.
%%%%%%%%%%%%%%%%%%%%%%%%%%%%%%%%%%%%%%%%%%%%%%%
\begin{figure}
%\vspace{0.6cm}
\begin{center}
\includegraphics[width=0.9\columnwidth]{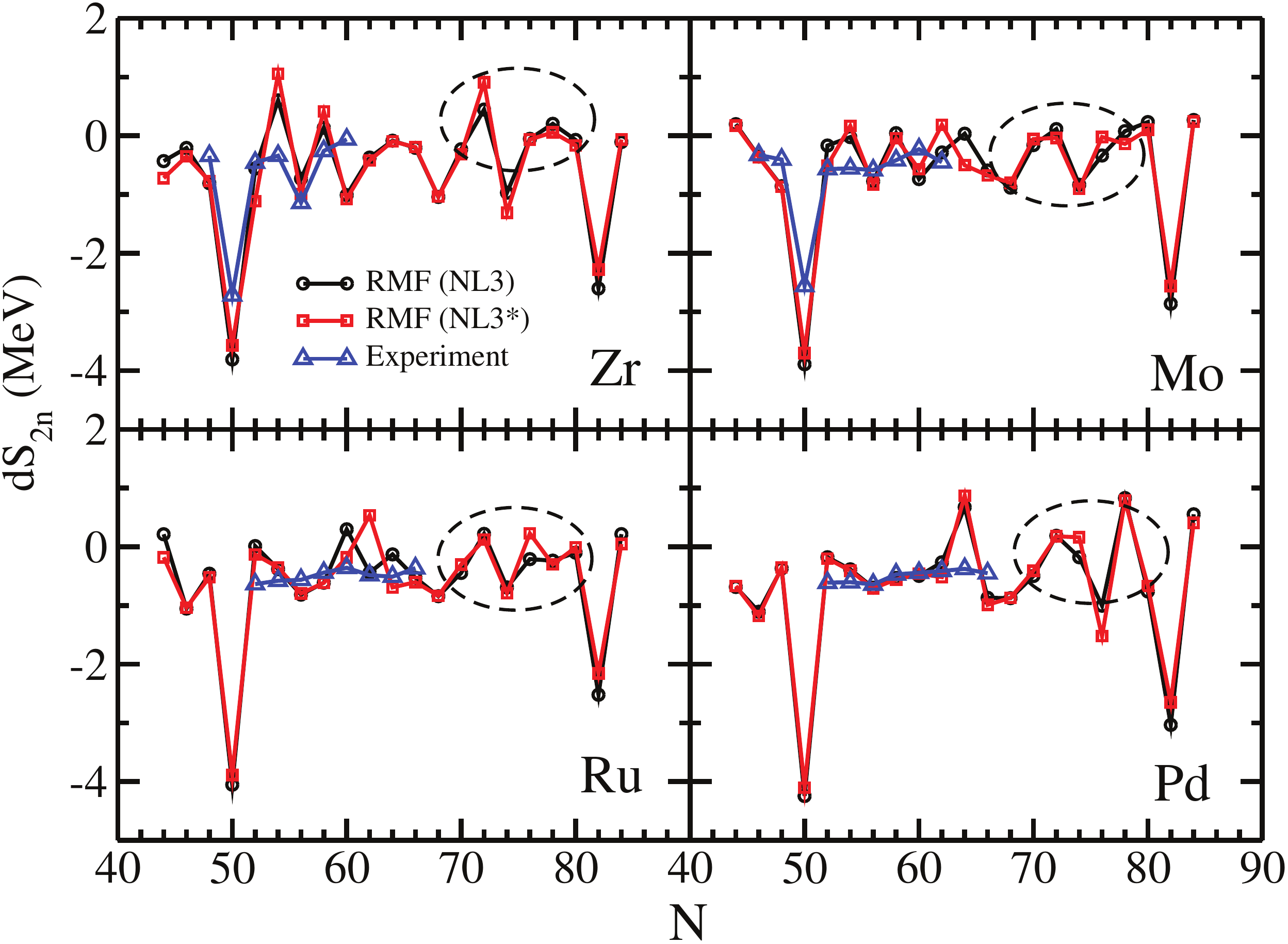}
\caption{(Color online) The differential variation of the two neutron
separation energy $dS_{2n}$ as a function of neutron number from RMF
theory with $NL3$ and $NL3^*$ force parameter for $^{82-126}$Zr,
$^{84-128}$Mo, $^{86-130}$Ru and $^{88-132}$Pd nuclei are compared with
the experimental data \cite{audi12}.
}
\end{center}
\label{Fig. 1}
\end{figure}
%%%%%%%%%%%%%%%%%%%%%%%%%%%%%%%%%%%%%%%%%%%%%%

\subsection{Differential variation of two neutron separation energy}
The differential variation of the two neutron separation energy ($S_{2n}$) with respect 
to the neutron number ($N$) i.e. $dS_{2n} (N,Z)$ is defined as
\begin{eqnarray}
dS_{2n}(Z,N)=\frac{S_{2n}(Z,N+2)-S_{2n}(Z,N)}{2}, 
\end{eqnarray}
The $dS_{2n} (N,Z)$ is one of the key quantity to explore the rate of change of separation 
energy with respect to the neutron number in an isotopic chain. Here, we are calculated 
the $dS_{2n} (N,Z)$ for NL3 $\&$ NL3$^*$ force parameter. Further, we have also estimated 
the $dS_{2n}$ (N,Z) energy from the experimental $S_{2n}$ energy. In Fig. 4, we are compared 
the experimental values with our calculation for Zr, Mo, Ru and Pd isotopes. In 
general, the large sharp deep fall in the $dS_{2n}$ over an isotopic chain shows the 
signature of neutron shell closure. In other word, this deviation in the general trend 
may disclose some additional nuclear structure features. From the figure, we observed the 
same characteristics for all $Z$=38-46. 
%%%%%%%%%%%%%%%%%%%%%%%%%%%%%%%%%%%%%%%%%%%%5
\begin{figure}
%\vspace{0.6cm}
\begin{center}
\includegraphics[width=1.0\columnwidth]{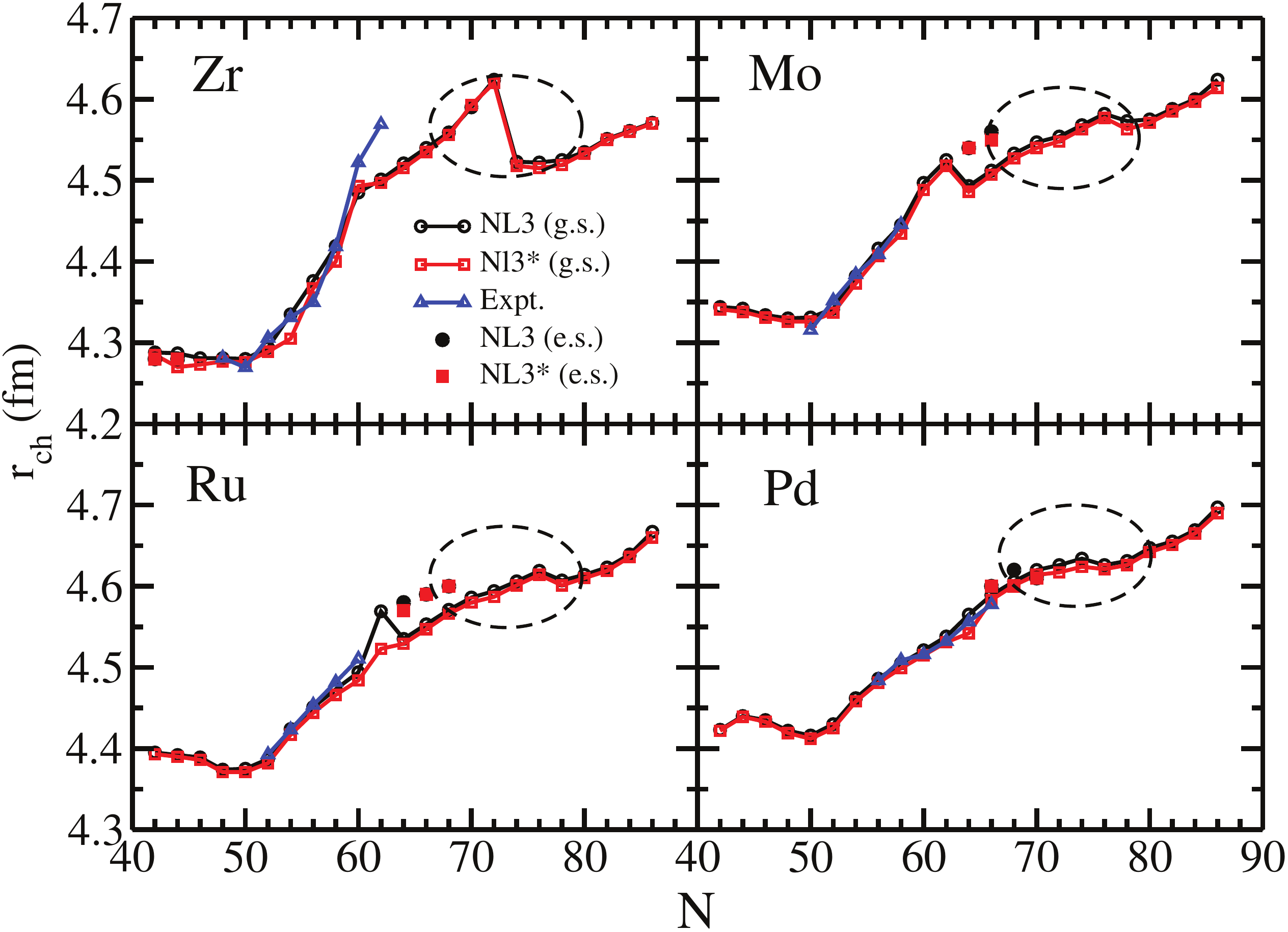}
\caption{(Color online) The root-mean-square charge distribution $r_{ch}$
of $^{82-126}$Zr, $^{84-128}$Mo, $^{86-130}$Ru and $^{88-132}$Pd nuclei from
RMF theory with $NL3$ and $NL3^*$ force parameter are compared with the
experimental data \cite{angeli13}.
}
\end{center}
\label{Fig. 1}
\end{figure}
%%%%%%%%%%%%%%%%%%%%%%%%%%%%%%%%%%%%%%%

\subsection{The root-mean-square charge distributions}
The root mean square ($rms$) matter radius from relativistic mean field theory can be 
expressed as:
\begin{eqnarray}
\langle r_m^2 \rangle = {1\over{A}}\int\rho(r_{\perp},z) r^2d\tau,
\end{eqnarray}
where $A$ is the mass number and $\rho(r_{\perp},z)$ is the axially deformed density. 
The $rms$ charge radius can be calculated from the $rms$ proton radius $\langle r_p^2 
\rangle$ with simple algebraic relation, 
\begin{eqnarray}
\langle r_{ch}^2 \rangle = \langle r_p^2 \rangle + 0.64.
\end{eqnarray}
From the theoretical point of view, the {\it macroscopic-microscopic} models 
\cite{wang01} and $microscopic$ mean-field formulations using effective interactions 
are most sophisticated approaches to determine the $rms$ charge radius in comparison 
with experimental data \cite{angeli13}. In this present work, we have shown the 
variations or fluctuations of the charge radii on the top of a fairly smooth average 
behavior in an isotopic chain. The results from RMF approaches for $NL3$ and $NL3^*$
parameters along with the available experimental data are shown in Fig. 5. From the 
figure it is clear that the obtained radii from RMF for $^{82-126}$Zr, $^{84-128}$Mo, 
$^{86-130}$Ru and $^{88-132}$Pd follows closely the experimental data \cite{angeli13}. 
For most of the nuclei, the experimental values are unavailable, the RMF prediction 
are made for the charge radius of such a nucleus that awaits experimental confirmation. 
The circle, square and triangle symbols indicate the ground state states data for NL3, 
NL3* and experiment, respectively. Further, the solid circle and solid square symbols 
indicate the shapes correspond to the first intrinsic excited states obtained from NL3 
and NL3* force. From the figure, one can observe the smooth behavior for lighter 
isotopes, then there is a small fall in the charge radii for Zr, Mo, and Pd at about 
$N$=62, 64, 72 and 74. This fall corresponds to the transition from the prolate to the 
oblate and vice versa. But the magnitude for both the states are different, i.e., the 
oblate deformation is at $\beta_2 \sim$ $-$ 0.2 while the prolate one appears with 
$\beta_2 \sim$ 0.4. In case of $Pb$ isotope, the change is the radii only at one place 
i.e. at $N$=74. Further, one can notice that the tiny change in the calculation can 
lead to the first intrinsic excited state as a ground state (see Fig. 2). In other word, 
we can practically degenerate the ground state binding energy for the deformation 
corresponding to the first intrinsic excited state. Thus, the inconsistency in the 
$r_c^2$ could be explained in terms of configuration mixing i.e the actual ground 
state is not only the spherical configuration but also from the neighbor deformed 
intrinsic excited states.
%%%%%%%%%%%%%%%%%%%%%%
\begin{figure}
%\vspace{0.6cm}
\begin{center}
\includegraphics[width=1.0\columnwidth]{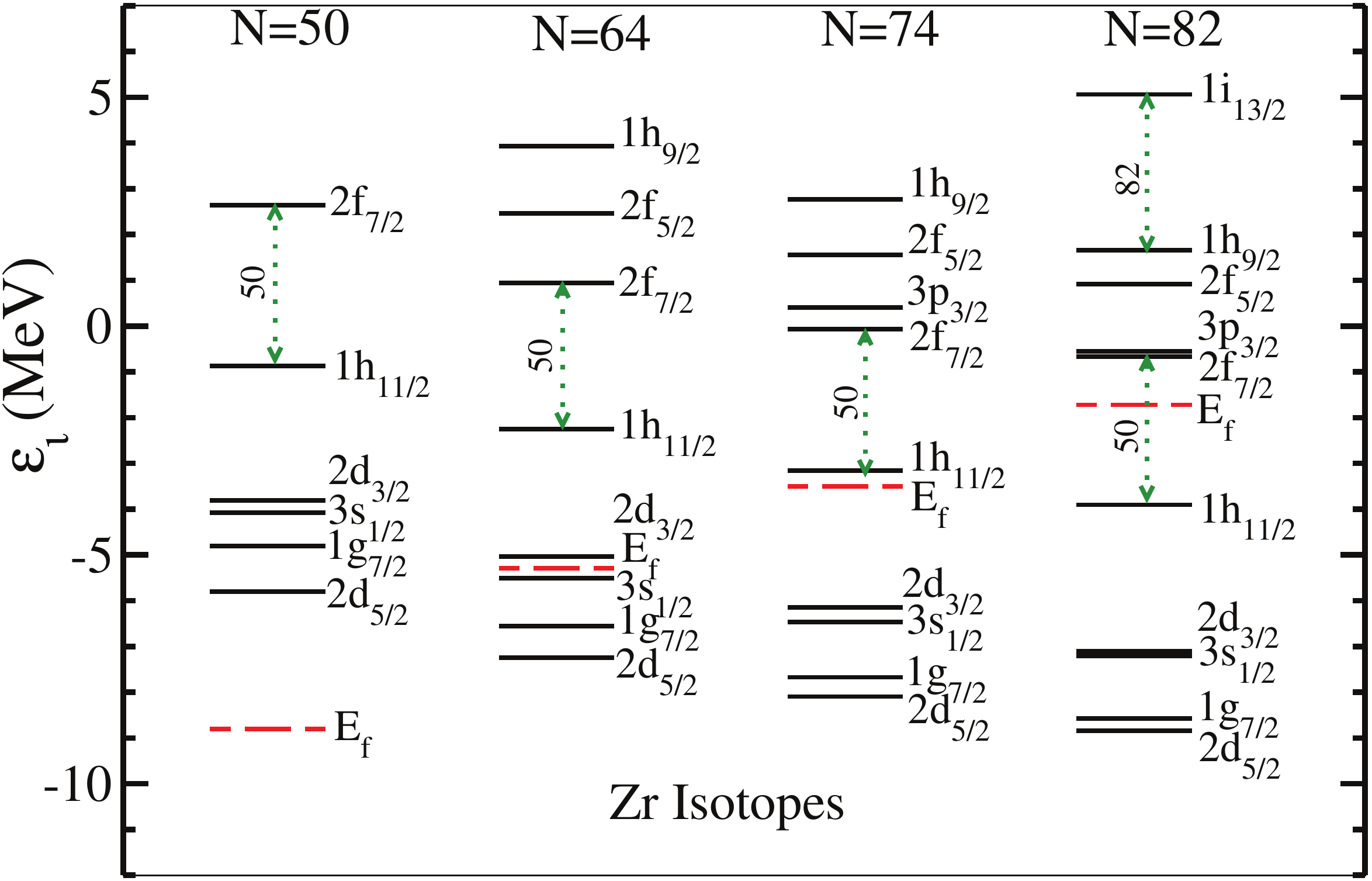}
\caption{(Color online) The single particle energy levels $\epsilon_i$ for 
$^{90}$Zr, $^{104}$Zr, $^{114}$Zr and $^{122}$Zr from RMF model with $NL3$ 
force parameter}
\end{center}
\label{Fig. 1}
\end{figure}
%%%%%%%%%%%%%%%%%%%%%%%%%%%5
\subsection{Single particle energy levels}
The above analysis, we found some signatures of shell closures at N=  
82 for all these isotopic chains. As a further confirmatory test, the 
single-particle energy levels for neutrons in isotopic chains are 
examined. The obtained single particle  levels $\epsilon_i$ for even 
isotopes of Zr near Fermi level are shown in Fig. 6 for NL3 force as an 
ideal case. However, we have obtained similar results for all isotopic 
chain from NL3 and NL3* force parameter. We observed the large gap at N 
= 82, near the drip-line region. In general, the spin-orbit splitting of 
the levels are scale down for neutron-rich nuclei, but $1h_{11/2}$ level 
(at N=82) is higher in the Zr nuclei studied.  Quantitatively, in $^{122}$Zr, 
the $\Delta \epsilon_i = \epsilon_i (1i_{13/2}) −  \epsilon_i(1h_{9/2})$ 
at N = 82 is 4.5 MeV, which is a considerably large value compare to the 
neighbor splitting. Almost identical behavior is noticed for the isotopic 
chain of Mo, Ru and Pd nucleus at N=82, irrespective of force parameter 
used. Such a rearrangement of the single-particle orbitals at N=82, well 
accepted the shell closure at N=82 for the considered transitional nuclei.
%%%%%%%%%%%%%%%%%%%%%%%%%%%%%%%%%%%%%%%%%%%%5
\begin{figure}
%\vspace{-2.0cm}
\begin{center}
\includegraphics[width=1.0\columnwidth]{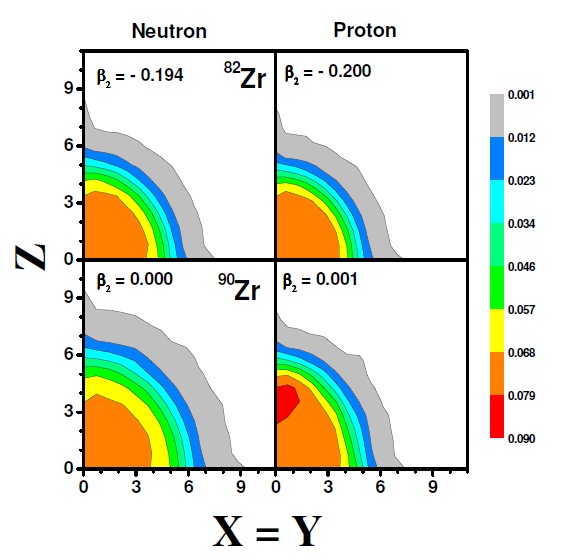}
%\vspace{-2.0cm}
\caption{(Color online) The contour plot of the axially deformed ground 
state density distribution of proton and neutron for $^{82}$Zr and $^{90}$Zr.}
\end{center}
\label{Fig. 1}
\end{figure}

\begin{figure}
%\vspace{-2.0cm}
\begin{center}
\includegraphics[width=1.0\columnwidth]{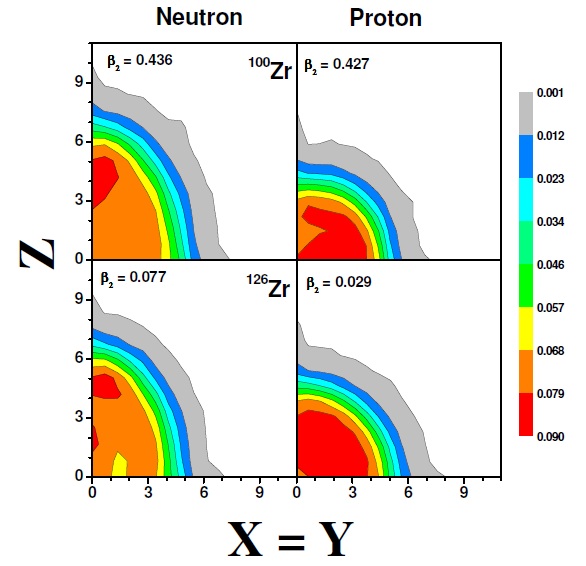}
%\vspace{-2.0cm}
\caption{(Color online) The contour plot of the axially deformed ground 
state density distribution of proton and neutron for $^{100}$Zr and $^{126}$Zr.}
\end{center}
\label{Fig. 1}
\end{figure}

%%%%%%%%%%%%%%%%%%%%%%%%%%%%%%%%%%%%%%%

\subsection{The contour plot of the axially deformed densities}
In the above figures and tables, we have shown the results some of the 
structural observables such as binding energy, quadrupole deformation, 
root-mean-square radius, separation energy, differential separation 
energy and single-particle energy levels in comparison with the 
experimental data \cite{audi12,angeli13}, wherever available. Here, we 
have focused on the ground of $^{82-126}$Zr, $^{84-128}$Mo, $^{86-130}$Ru 
and $^{88-132}$Pd nuclei along with few selectively excited states. Based 
on these structural observables, we found some significant signature of 
the shell closure at $N$ = 82 (drip-line region) in the isotopic chains. 
Further, the abnormal change in the $S_{2n}$ and $dS_{2n}$ in the isotopic 
chain of Zr, Mo and Ru nucleus suggest  a shape co-existence at $N 
\sim$ 64 and 74. This divergence over an isotopic chain can be cut down 
by taking the dynamical correlations beyond mean-field 
\cite{bender06,bender08,dela10}.

To get a complete picture into the reason behind such discrepancy over 
the isotopic chain, we have shown the contour plot of the axially deformed 
density of proton and neutron of these nuclei. In Fig. 7-8, we have 
displayed the distribution of Zr isotopes for $N$=42, 50, 60 and 82 
as representative cases. All the isotopes of Mo, Ru, and Pd also 
showing similar behavior as Zr as shown in Fig. 7-8. From the figure, 
one can clear identify the spherical, oblate, prolate shapes corresponding 
to their $\beta_2$ values as the local minima in the PECs. Similar 
calculations can also be found in Ref. \cite{web12,plb12}. In these 
figures, we can see that the transition from oblate to prolate at $N$=42, 
then change to the spherical structure at $N$= 50 and further changing 
the deformations to prolate one. Even the proton number is fixed in the 
isotopic chain, still we found a little change in the density distribution 
due to the influence of excess neutron number. Following the color code, 
the red and light gray color corresponding to the high density ($\sim$ 
0.09 fm$^{-3}$) and low density ($\sim$ 0.001fm$^{-3}$), respectively. 
More inspection on the figures shows that the central density of the 
proton increases as compared to the neutron with respect to the neutron 
number. In this region, few isotopes of Mo (for $^{116-118}$Mo) are the 
triaxial shape in their ground state, which is very close to the axial 
solutions \cite{plb12}. In other word, the location of minima for a 
triaxial solution for these isotopes of Mo are almost same as the minima 
appear for an axial prolate axial solution. Hence, we have used the simple 
axial deformed calculation, which is adequate for the a qualitative 
descriptions of structural observables in this mass region.

\section{Summary and Conclusions}
We have used self-consistent relativistic mean field theory with most popular 
$NL3$ and recent $NL3^*$ force parameters to study the structural evolution 
in transition nuclei. The conjecture has been made from the binding energy, 
neutron separation energies, differential variation of separation energy, 
the root-mean-square charge radii and the single particle energy levels of 
these nuclei. In this present calculations we have shown that Zr, Mo and 
Ru isotopes undergo a transition from oblate to prolate shapes at $N \sim$ 
64 and 74. But, in case of Pd follows a smooth pattern through out the 
isotopic chain. We have also shown the dependence of nuclear charge radii 
on deformation also play an crucial role on their structural transition. 
Further, we have also observed a large shell gap at $N$ = 82 near drip-line 
region, almost same in magnitude at $N$= 50 for these considered nuclei, which 
is a well-known feature for mean-field calculation. We have also demonstrated 
the efficiency of RMF theory calculations to reproduce those features and 
therefore to make predictions in unexplored regions.

\section*{Acknowledgments}
The author thanks to S. K. Patra, Institute of Physics, Bhubaneswar and Shan-Gui 
Zhou, Institute of Theoretical Physics, Chinese Academy of Sciences for various 
discussion regarding this work. Further, the author thankful to Cheng-Jun 
Xia, Jie Zhao, Bing Wang, Mung Xu and Zhen-Hua Zhang for careful reading 
of the manuscript.

\end{document}